\documentclass[twocolumn]{jpsj3}
\usepackage{bm,amssymb,amsmath}	
\usepackage{graphicx}   
\usepackage{txfonts}
\newcommand{\simg}{\stackrel{>}{_\sim}}

\title{
\begin{center}
A High-$T_{c}$ Mechanism of Iron Pnictide Superconductivity due to Cooperation of Ferro-orbital and Antiferromagnetic Fluctuations
\end{center}
}

\author{Takemi {\sc Yamada}\thanks{E-mail address: takemi@phys.sc.niigata-u.ac.jp}, Jun {\sc Ishizuka} and Yoshiaki {\sc \=Ono}}
\inst{Department of Physics, Niigata University, Ikarashi, Niigata, 950-2181, Japan}

\abst{
The electronic states and superconductivity in iron pnictides are studied on the basis of the 16 band $d$-$p$ model 
which includes both the onsite Coulomb interaction between Fe $d$ electrons 
and the intersite one between Fe $d$ and pnictogen $p$ electrons. 
The model well accounts for experimentally observed two fluctuations: 
the $d$-$d$ interaction-enhanced antiferromagnetic (AFM) fluctuation and 
the $d$-$p$ interaction-enhanced ferro-orbital (FO) fluctuation responsible for the $C_{66}$ elastic softening. 
The AFM fluctuation induces the repulsive pairing interaction for $\bm{q}\sim \bm{Q}_{\rm AF}$ 
while the FO does the attractive one for $\bm{q}\sim \bm{0}$ resulting in the $s_{\pm}$-wave superconductivity 
where the two fluctuations cooperatively enhance the superconducting transition temperature $T_{c}$ 
without any competition by virtue of the $\bm{q}$-space segregation. 
}

\begin{document}
\maketitle
Since the discovery of superconductivity with high transition temperature $T_c$ in LaFeAsO$_{1-x}$F$_x$\cite{JACS.130.3296}, 
the pairing mechanisms of the iron pnictide superconductors have attracted much attention\cite{AdvPhys.59.803,RepProgPhys.74.124508}. 
Two significant fluctuations: the stripe-type antiferromagnetic (AFM) fluctuation\cite{AdvPhys.59.803} 
which diverges towards the AFM transition and the ferro-orbital (FO) fluctuation corresponding to the $O_{xy}$ ferro-quadrupole one 
which is responsible for the softening of the elastic constant $C_{66}$\cite{PhysRevLett.105.157003,JPSJ.80.073702,JPSJ.81.024604} 
and diverges towards the tetragonal-orthorhombic structural transition\cite{C66}, 
have been discussed as key ingredients for the pairing mechanisms. 
Theoretically, the AFM fluctuation for $\bm{q}\sim \bm{Q}_{\rm AF}$ corresponding 
to the nesting wave vector between electron and hole Fermi surfaces (FSs) 
was found to be enhanced by the onsite Coulomb interaction between Fe $d$ electrons 
and to induce the repulsive pairing interaction for $\bm{q}\sim \bm{Q}_{\rm AF}$ resulting in the $s_{\pm}$-wave superconductivity 
where the gap function changes its sign  between the electron and the hole FSs\cite{PhysRevLett.101.057003,PhysRevLett.101.087004}.

On the other hand, the FO fluctuation for $\bm{q}\sim \bm{0}$ was found to be enhanced by the electron-phonon interaction\cite{JPSJ.79.123707} 
and/or the mode-coupling\cite{PhysRevB.84.024528,PhysRevLett.109.137001}, 
where the antiferro-orbital (AFO) fluctuation for $\bm{q}\sim \bm{Q}_{\rm AF}$ was also enhanced 
due to the nesting as similar to the AFM fluctuation\cite{JPSJ.79.123707,PhysRevB.84.024528,PhysRevLett.109.137001,PhysRevLett.104.157001,PhysRevB.82.064518}. 
When the attractive pairing interaction mediated by the AFO fluctuation overcomes the repulsive one 
by the AFM fluctuation for the same wave vector $\bm{q}\sim \bm{Q}_{\rm AF}$, 
the $s_{++}$-wave superconductivity without the sign change of the gap function was found to be realized 
with the help of the attractive one for $\bm{q}\sim \bm{0}$ by the FO fluctuation\cite{JPSJ.79.123707,PhysRevB.84.024528,PhysRevLett.109.137001,PhysRevLett.104.157001,PhysRevB.82.064518}. 
At the moment it is not clear which fluctuation is dominant for $\bm{q}\sim \bm{Q}_{\rm AF}$ 
that is crucial in determining whether the $s_{\pm}$- or the $s_{++}$-wave takes place, 
since the AFO fluctuation has not been explicitly observed in experiments so far\cite{phonon}. 
In either case, the AFM and the AFO fluctuations compete with each other for the pairing interaction resulting in suppression of $T_c$ 
as compared to the case with either fluctuation alone.  

In this letter, we propose another mechanism of the FO fluctuation enhancement 
due to the intersite Coulomb interaction between Fe $d$ and pnictogen $p$ electrons. 
A recent experiment has actually provided evidence for strong coupling of Fe and pnictogen orbital polarizations (OPs)\cite{arXiv.1401.3706}. 
Then, we employ the 16 band $d$-$p$ model which explicitly includes Fe $3d$ and As $4p$ orbitals reproducing the band structure of LaFeAsO 
and has been extensively studied focusing on the effects of the $d$-$d$ interaction\cite{JPSJ.77.123701,PhysRevB.81.054518} 
and/or the electron-phonon interaction\cite{JPSJ.79.123707,PhysRevB.82.064518,SSC.152.701}. The effect of the $d$-$p$ interaction 
has also been investigated and found to enhance the charge fluctuation 
which mediates the $s_{\pm}$- or the $s_{++}$-wave pairing depending on the parameters\cite{Yanagi.PhD2010,PhysRevB.84.140505}. 
However, the interaction between the Fe and As OPs depending on relative direction of $d$ and $p$ orbitals has not been considered there. 
We find that the $d$-$p$ OP interaction enhances the FO fluctuation responsible for the $C_{66}$ softening 
without enhancing the AFO one resulting in the $s_{\pm}$-wave superconductivity in collaboration with the AFM fluctuation enhanced by the $d$-$d$ interaction. 
In this case, the experimentally observed two fluctuations cooperatively enhance $T_{c}$ without any competition by virtue of the $\bm{q}$-space segregation.

\begin{figure}[t]
\begin{center}
\includegraphics[width=6.8cm]{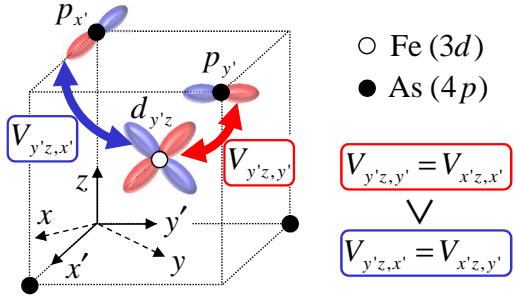}
\vspace{-0.1cm}
\caption{(Color online) 
The origin of the $d$-$p$ orbital polarization interaction $V'=(V_{x'z,x'}-V_{x'z,y'})/2$ 
due to the orbital dependence of Coulomb integrals between neighboring Fe and As sites with $V_{y'z,y'}(=V_{x'z,x'})>V_{y'z,x'}(=V_{x'z,y'})$, 
where $x'$, $y'$ ($x$, $y$) refer to the direction along the nearest (second nearest) Fe-Fe bonds. 
}
\label{Fig1}
\end{center}
\end{figure}

Our 16 band $d$-$p$ model consists of 16 orbitals in each unit cell: five $3d$ orbitals of two Fe atoms and three $4p$ orbitals of two As atoms and is given by 
\begin{align}
H=H_0+H_{\rm int}^{dd}+H_{\rm int}^{pp}+H_{\rm int}^{dp}, 
\label{eq:H}
\end{align}
where $H_0$ is the non-interacting tight-binding Hamiltonian derived 
so as to reproduce the band structure of LaFeAsO\cite{JPSJ.77.123701,PhysRevB.81.054518} 
and $H_{\rm int}^{dd}$, $H_{\rm int}^{pp}$ and $H_{\rm int}^{dp}$ represent the Coulomb interaction between the onsite Fe $d$ electrons, 
the onsite As $p$ electrons and the intersite Fe $d$ and neighboring As $p$ electrons, respectively. 
From the first-principles downfolding scheme, Miyake {\it et al.}\cite{JPSJ.79.044705} revealed that 
the Coulomb and the exchange integrals $U_{ll'}$ and $J_{ll'}$ in $H_{\rm int}^{dd}$ are orbital ($l,l'$) dependent 
and the average of $U_{ll}$ is $U_d=4.2$eV for LaFeAsO, while $H_{\rm int}^{pp}$ and $H_{\rm int}^{dp}$ are also large: $U_p=2.5$eV (LaFePO) and $U_{pd}=1.2$eV (LaFeAsO). 
As the explicit orbital dependence of $H_{\rm int}^{pp}$ and $H_{\rm int}^{dp}$ is not obtained so far, 
we set $U_{ll'}=U_p$ and $J_{ll'}=0$ in $H_{\rm int}^{pp}$ for simplicity, and we assume that 
\begin{equation}
H_{\rm int}^{dp}=V\sum_{\langle i,j\rangle}\hat{n}_{i}^{d}\hat{n}_{j}^{p}
+V'\sum_{\langle i,j\rangle}(\hat{n}_{ix'z}^{d}-\hat{n}_{iy'z}^{d})(\hat{n}_{jx'}^{p}-\hat{n}_{jy'}^{p}),
\label{eq:Hdp}
\end{equation}
where $\hat{n}_{il}^{d}$ ($\hat{n}_{jm}^{p}$) is the $d$ ($p$) electron number operator for orbital $l$ ($m$), 
$\hat{n}_{i}^{d}=\sum_l\hat{n}_{il}^{d}$ ($\hat{n}_{j}^{p}=\sum_m\hat{n}_{jm}^{p}$) and $\langle i,j\rangle$ are nearest-neighbor Fe and As sites. 
In eq. (\ref{eq:Hdp}), 
$V$ is the $d$-$p$ charge transfer interaction which was found to enhance the charge fluctuation 
as originally discussed for the cuprate superconductors\cite{PhysRevLett.63.2602,JPSJ.60.2269} 
and also for the iron pnictides\cite{Yanagi.PhD2010,PhysRevB.84.140505}, 
and $V'=(V_{x'z,x'}-V_{x'z,y'})/2$ is the $d$-$p$ OP interaction derived from the orbital dependence of the Coulomb integrals 
between Fe $d_l$ and As $p_m$ orbitals $V_{l,m}$ for $l=d_{x'z}$ or $d_{y'z}$ and $m=p_{x'}$ or $p_{y'}$ as shown in Fig. \ref{Fig1}, 
where $x'$, $y'$ ($x$, $y$) refer to the direction along the nearest (second nearest) Fe-Fe bonds. 
A rough estimation with the use of hydrogenlike atomic wave functions yields a considerably large value of $V'$ with several tenth percent of $V$. 
Here, we focus only on the $d_{x'z}$-$d_{y'z}$ type OP interaction crucial for the FO fluctuation responsible for the $C_{66}$ softening 
but the effects of the other types of OP interaction will be discussed later. 
We also neglect the interaction between Fe charge and As OP 
which was found to enhance the charge fluctuations for $\bm{q}\ne\bm{0}$\cite{PhysRevB.84.140505} but is almost irrelevant for the FO fluctuation. 

Now, we investigate the Hamiltonian eq. (\ref{eq:H}) within the random phase approximation (RPA), 
where the spin and charge-orbital susceptibilities are given in the $68\times68$ matrix representation as\cite{Yanagi.PhD2010} 
\begin{align}
\hat{\chi}^{s}(q)&=\left[\hat{1} -\hat{\chi}^0(q)\hat{\Gamma}^{s}\right]^{-1} \hat{\chi}_0(q), 
\label{eq:chis}\\
\hat{\chi}^{c}(q)&=\left[\hat{1} +\hat{\chi}^0(q)\hat{\Gamma}^{c}(\bm{q})\right]^{-1} \hat{\chi}_0(q) 
\label{eq:chic}
\end{align}
with the noninteracting susceptibility ${\chi}^0_{\mu_1 \mu_2 \mu_3 \mu_4}(q)=-\frac{T}{N}\sum_{k}{G}_{\mu_3 \mu_1}(k+q){G}_{\mu_2 \mu_4}(k)$, 
where $\mu$ represents Fe $d$ or As $p$ orbitals 
and $\hat{G}(k)=[i\varepsilon_m\hat{1}-\hat{H}_0(\bm{k})]^{-1}$ is the noninteracting Green's function ($16\times16$ matrix); 
$k=(\bm{k},i\varepsilon_m)$ and $q=(\bm{q},i\omega_n)$ with the wave vectors $\bm{k}, \bm{q}$ 
and the Matsubara frequencies $\varepsilon_{m}=2(m+1)\pi T$, 
$\omega_n=2n\pi T$, and $\hat{1}$ is the unit matrix in $\mu$ basis. 
In eqs. (\ref{eq:chis}) and (\ref{eq:chic}), $\hat{\Gamma}^{s(c)}$ is the spin (charge-orbital) vertex ($68\times68$ matrix) 
in which the nonzero elements are as follows: 
$\Gamma_{llll}^{s(c)}=U_{ll}~(U_{ll})$, 
$\Gamma_{ll'll'}^{s(c)}=U_{ll'}~(-U_{ll'}+2J_{ll'})$, 
$\Gamma_{lll'l'}^{s(c)}=J_{ll'}~(2U_{ll'}-J_{ll'})$, 
$\Gamma_{ll'l'l}^{s(c)}=J_{ll'}~(J_{ll'})$ 
in the $25\times25$ $d$-$d$ submatrix for each Fe atom: Fe$^1$, Fe$^2$, and   
$\Gamma_{mmmm}^{s}=\Gamma_{mm'mm'}^{s}=\Gamma_{mmmm}^{c}=-\Gamma_{mm'mm'}^{c}=U_{p}$, 
$\Gamma_{mmm'm'}^{c}=2U_{p}$ 
in the $9\times9$ $p$-$p$ submatrix for each As atom: As$^1$, As$^2$, 
where $l(\ne l')=d_{x^2-y^2}$, $d_{3z^2-r^2}$, $d_{xz}$, $d_{yz}$, $d_{xy}$ 
and $m(\ne m')=p_x$, $p_y$, $p_z$.  
The nonzero elements in the $50\times18$ $d$-$p$ submatrix are: 
$\Gamma_{llmm}^{c}(\bm{q})$=$2V\phi_{\alpha\beta}(\bm{q})$ for all $l=d_l$ and $m=p_m$,  
$\Gamma_{ll'mm'}^{c}(\bm{q})$=$2V'\phi_{\alpha\beta}(\bm{q})$ 
for $l(\ne l')=d_{xz}$, $d_{yz}$ and $m(\ne m')=p_{x}$, $p_{y}$  
with the $\bm{q}$ dependent factor $\phi_{\alpha\beta}(\bm{q})=1+e^{iq_{\nu}}$ due to intersite Fe-As contributions, 
where $q_{\nu}=-(+)q_x$ for $(\alpha, \beta)=(\mbox{Fe}^{1(2)}, \mbox{As}^{1(2)})$
and $q_{\nu}=-(+)q_y$  for $(\alpha, \beta)=(\mbox{Fe}^{1(2)}, \mbox{As}^{2(1)})$. 
Here, we note that the longitudinal $d_{x'z}$-$d_{y'z}$ and $p_{x'}$-$p_{y'}$ polarizations 
coupled via $V'$ with each other (see eq. (\ref{eq:Hdp})) are transformed into the transverse $d_{xz}$-$d_{yz}$ and $p_{x}$-$p_{y}$ ones 
by $45^\circ$ rotation as shown in Fig. \ref{Fig1}, respectively.

When the largest eigenvalue $\alpha_{s(c)}(\bm{q})$, called the spin (charge-orbital) Stoner factor, 
of the matrix $(-)\hat{\chi}^0(q)\hat{\Gamma}^{s(c)}$ in eq. (\ref{eq:chis}) (eq. (\ref{eq:chic})) for a wave vector $\bm{q}$ with $i\omega_n=0$ reaches unity, 
the instability towards the magnetic (charge-orbital) order with the corresponding $\bm{q}$ takes place. 
To examine the superconductivity, we solve the linearized Eliashberg equation 
\begin{align}
\lambda \Delta_{\mu\mu'}(k)=
&-\frac{T}{N}\!\sum_{k'}\!\sum_{\mu_1\mu_2\mu_3\mu_4}V_{\mu\mu_1\mu_2\mu'}(k-k') \nonumber \\
&\times G_{\mu_3\mu_1}(-k')\Delta_{\mu_3\mu_4}(k')G_{\mu_4\mu_2}(k'), 
\label{eq:gap} 
\end{align}
and obtain the superconducting gap function $\hat{\Delta}(k)$ ($16\times16$ matrix) with the eigenvalue $\lambda$ 
which becomes unity at the superconducting transition temperature $T_c$, 
where the effective pairing interaction for the spin-singlet state is given in the $68\times 68$ matrix representation as 
\begin{align}
\hat{V}(q)
\!=\!
\frac{3}{2}\hat{\Gamma}^{s}\hat{\chi}^{s}(q)\hat{\Gamma}^{s}
\!-\!
\frac{1}{2}\hat{\Gamma}^{c}(\bm{q})\hat{\chi}^{c}(q)\hat{\Gamma}^{c}(\bm{q})
\!+\!
\frac{1}{2}\left(\!\hat{\Gamma}^{s}\!+\!\hat{\Gamma}^{c}(\bm{q})\!\right).
\label{eq:pair}
\end{align}
The RPA calculations are performed with 32$\times$32 $\bm{k}$-point meshes and 512 Matsubara frequencies for the temperature $T=0.02$eV 
and the number of electrons per unit cell $n=24.2$ corresponding to $10\%$ electron doping. 
We employ the $d$-$d$ interaction parameters $U_{ll'}$ and $J_{ll'}$ obtained in Ref.\cite{JPSJ.79.044705} 
by multiplying a reduction factor $f_d=0.37-0.40$ as done for the 5 orbital Hubbard model with $f=0.42$\cite{JPSJ.82.083702} 
since the RPA overestimates the magnetic order and fluctuation as explicitly shown in Ref.\cite{JPSJ.82.123712}. 
The $p$-$p$ interaction $U_p=2.5$eV is taken from Ref.\cite{JPSJ.79.044705}, 
and the $d$-$p$ charge transfer interaction $V=0.5$eV is assumed to be smaller than $U_{pd}=1.2$eV from Ref.\cite{JPSJ.79.044705} 
in order to avoid the instability towards the phase separation that occurs for $V\simg 0.57$eV\cite{CTI} 
but disappears with taking proper account of the long-range Coulomb interaction\cite{JPSJ.61.649} 
which suppresses the uniform charge fluctuation due to charge screening effect\cite{PhysRevB.84.140505} 
but not the FO fluctuation enhanced by the $d$-$p$ OP interaction $V'$ as mentioned below.

\begin{figure}[t]
\begin{center}
\includegraphics[width=7.2cm]{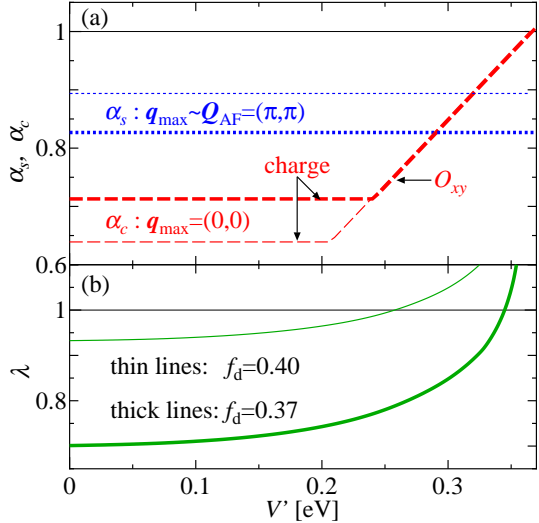}
\vspace{-0.3cm}
\caption{(Color online) The spin and charge-orbital Stoner factors $\alpha_{s}$ (dotted lines) and $\alpha_{c}$ (dashed lines) for ${\bm q}_{\rm max}$ (a), 
and the eigenvalue of the Eliashberg equation $\lambda$ (b) 
for the $d$-$d$ interaction reduction factor $f_d=0.37$ (thick lines) and $f_d=0.40$ (thin lines) as functions of the $d$-$p$ orbital polarization interaction $V'$. 
}
\label{Fig2}
\end{center}
\end{figure}

In Fig. \ref{Fig2} (a), we plot the spin (charge-orbital) Stoner factor $\alpha_{s(c)}$ for ${\bm q}={\bm q}_{\rm max}$ 
at which $\alpha_{s(c)}$ shows a maximum as a function of the $d$-$p$ OP interaction $V'$ for the $d$-$d$ interaction reduction factor $f_d=0.37$ and $0.40$. 
As $V'$ appears only in the $d_{xz}$-$d_{yz}$ off-diagonal elements of $\hat{\Gamma}^{c}(\bm{q})$ 
which takes a maximum value of $4V'$ at $\bm{q}=\bm{0}$ due to the $\bm{q}$ dependent factor $\phi_{\alpha\beta}(\bm{q})$, 
$V'$ enhances the transverse $d_{xz}$-$d_{yz}$ (longitudinal $d_{x'z}$-$d_{y'z}$) susceptibility 
which contributes to the $O_{xy}$ quadrupole one 
$\chi_{O_{xy}}({\bm q})=\sum_{l_1 l_2 l_3 l_4}[\hat{O}_{xy}]_{l_1 l_2}[\hat{O}_{xy}]_{l_4 l_3}{\chi}^c_{l_1 l_2 l_3 l_4}(\bm{q},0)$ 
for $\bm{q}\sim\bm{0}$ responsible for the $C_{66}$ softening\cite{C66}, 
while the charge susceptibility $\chi_c({\bm q})$ which is enhanced by $V$ for $\bm{q}\sim\bm{0}$\cite{Yanagi.PhD2010,PhysRevB.84.140505} is independent of $V'$. 
Therefore, $\chi_{O_{xy}}({\bm 0})$ monotonically increases with increasing $V'$ 
and dominates over $\chi_c({\bm 0})$ at a certain $V'$ at 
which $\alpha_{c}$ for ${\bm q}_{\rm max}={\bm 0}$ shows a kink as shown in Fig. \ref{Fig2} (a). 
Above the kink, $\alpha_{c}$ linearly increases with increasing $V'$ and finally reaches unity at a critical value $V'_c=0.365$eV 
where the instability towards the FO order with different occupations of $d_{x'z}$ and $d_{y'z}$ orbitals 
observed in the orthorhombic phase\cite{PhysRevLett.101.057010} takes place. 
We see from comparison between the results with $f_d=0.37$ and $0.40$ 
that the $d$-$d$ interaction enhances the stripe-type AFM susceptibility for $\bm{q}\sim \bm{Q}_{\rm AF}=(\pi,\pi)$, 
which is independent of $V$ and $V'$ included only in $\hat{\Gamma}^{c}(\bm{q})$, and suppresses $\chi_c({\bm 0})$ 
while keeping $\chi_{O_{xy}}({\bm 0})$ almost unchanged. 

\begin{figure}[t]
\begin{center}
\includegraphics[width=8.2cm]{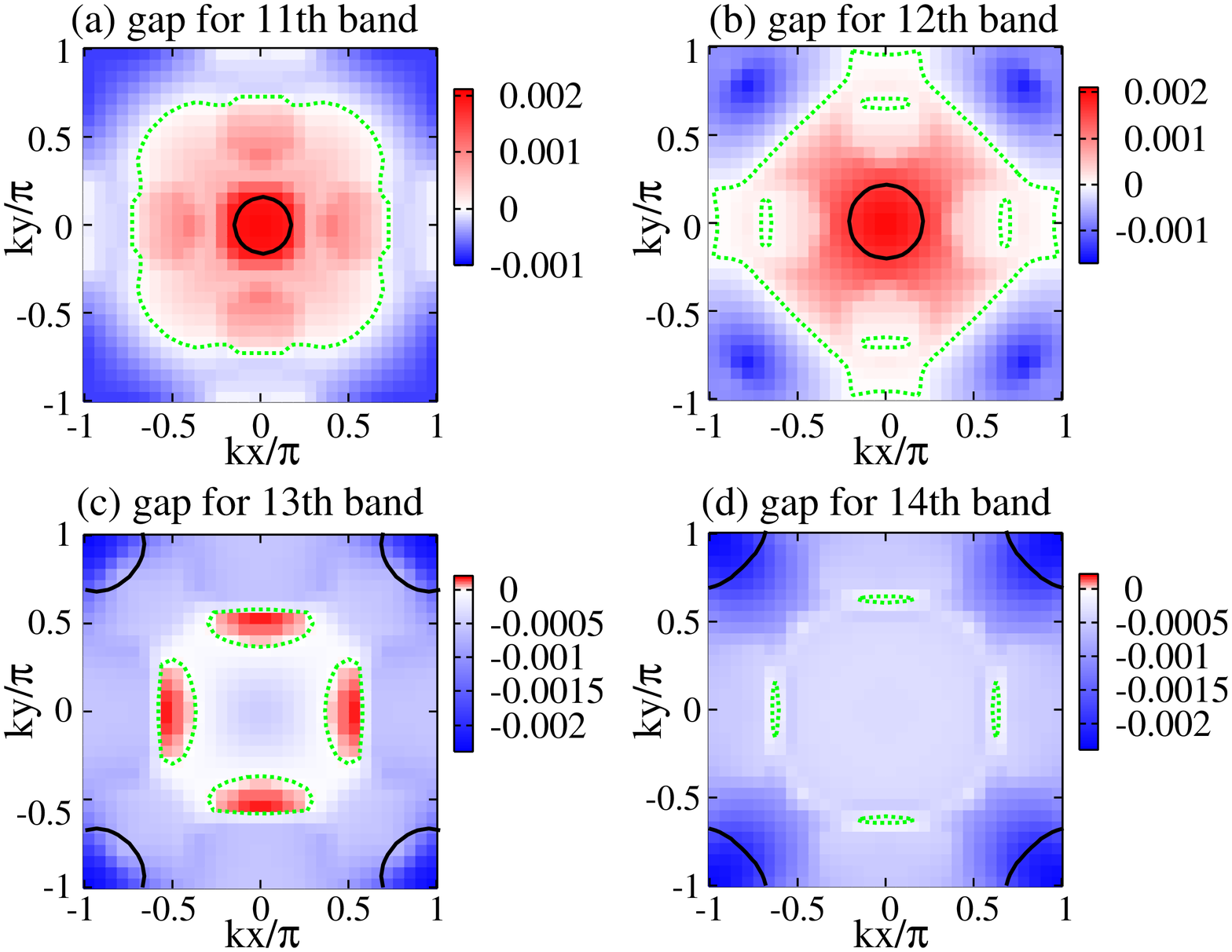}
\caption{(Color online) 
The band representation of the gap function $\hat{\Delta}(\bm{k},i\pi T)$ 
for the 11th (a) and 12th (b) (hole) bands and the 13th (c) and 14th (d) (electron) bands in the Brillouin zone 
corresponding to two FeAs per unit cell for $f_d=0.37$ and $V'=0.345$eV with $\lambda=1$, 
where black solid and green dotted lines represent the FSs and the nodes of the gap functions, respectively. 
}\label{Fig3}
\end{center}
\end{figure}

Fig. \ref{Fig2} (b) shows the eigenvalue $\lambda$ of the Eliashberg equation (\ref{eq:gap}) as a function of $V'$ for $f_d=0.37$ and $0.40$. 
For $f_d=0.37$ $(0.40)$, $\lambda$ reaches unity at $V'=0.345$ $(0.255)$eV 
where $\alpha_c=0.955$ $(\alpha_s=0.894)$ is larger than $\alpha_s=0.827$ $(\alpha_c=0.750)$ 
and then the FO (AFM) fluctuation gives a larger contribution to the superconductivity relative to the AFM (FO) one. 
In both cases,  $V'\sim 0.3$eV is a realistic value of the parameter as compared to the roughly estimated value mentioned before. 
In the FO fluctuation-dominated case with $f_d=0.37$, the gap function on the hole FSs (Figs. \ref{Fig3} (a) and (b)) 
and the electron FSs (Figs. \ref{Fig3} (c) and (d)) has almost the same absolute value without nodes 
and changes its sign between the hole and electron FSs, i.e., the fully gapped $s_{\pm}$-wave state 
which is obtained also for the AFM fluctuation-dominated case with $f_d=0.40$ (not shown). 
In the orbital representation, the $d$-$d$ component $\Delta_{ll'}^{dd}$ with $l=l'=d_{xz}$ or $d_{yz}$ has the largest value 
but the $d$-$p$ component $\Delta_{lm}^{dp}$ with $l=d_{xz}$ ($d_{yz}$) and $m=p_{y}$ ($p_x$) is also large $\sim 40\%$ of $\Delta_{ll}^{dd}$, 
while the $p$-$p$ component $\Delta_{mm'}^{pp}$ is small $\sim$ less than $1/10$ of  $\Delta_{ll}^{dd}$. 
Then, the $d$-$p$ correlation effects are important not only for the FO fluctuation enhancement inducing the pairing interaction 
but also for the superconducting state itself. 
To see the latter effect explicitly, we solve the Eliashberg equation (\ref{eq:gap}) in the absence of $\Delta_{lm}^{dp}$ 
and find that the obtained $\lambda$ is about $0.2$ smaller than that in the presence of  $\Delta_{lm}^{dp}$.

\begin{figure}[t]
\begin{center}
\includegraphics[width=8.0cm]{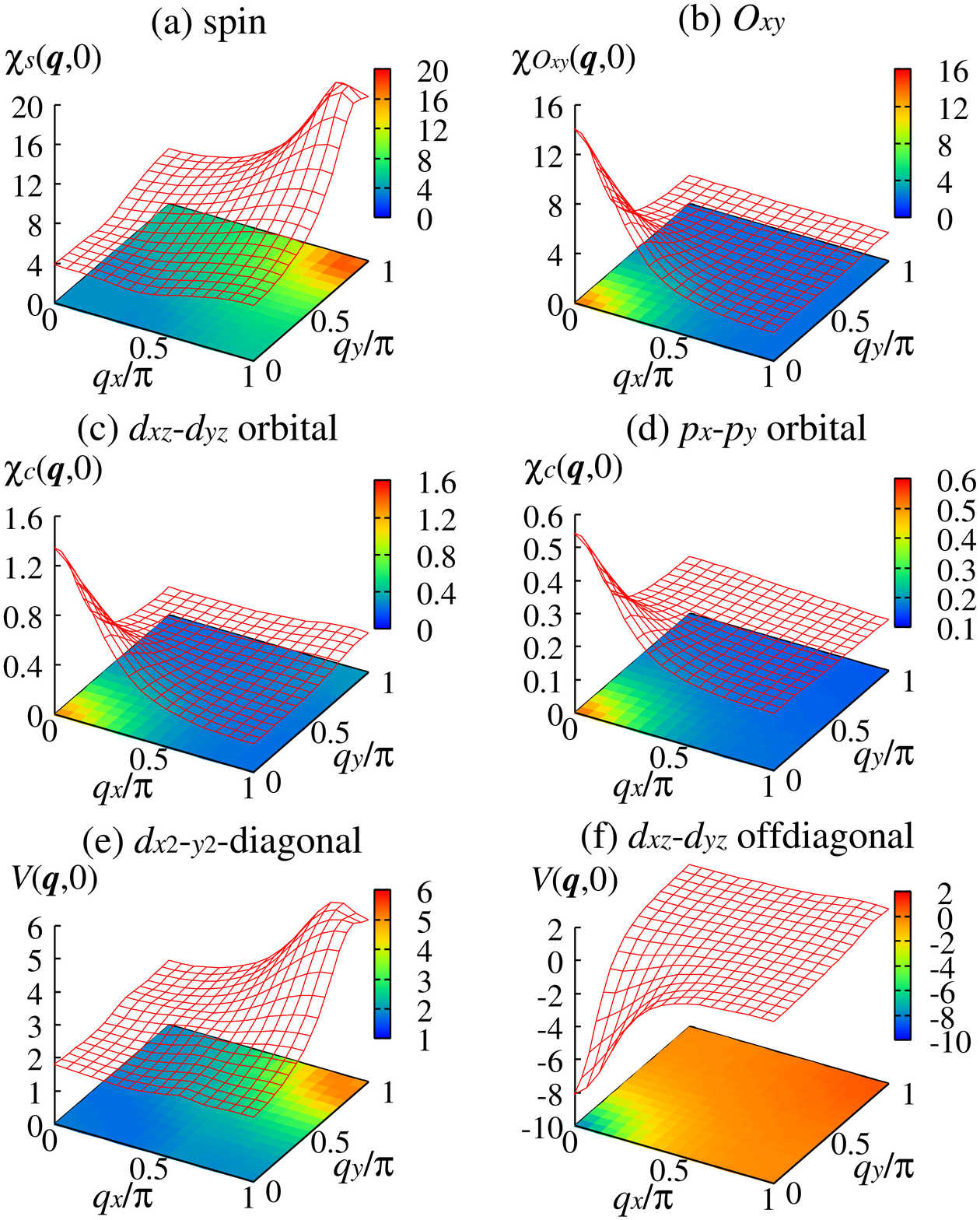}
\caption{(Color online) 
The susceptibilities for 
(a) spin $\chi_{s}$, 
(b) $O_{xy}$ quadrupole $\chi_{O_{xy}}$, 
(c) transverse $d_{xz}$-$d_{yz}$ orbital $\chi^{c}_{xz,yz,xz,yz}$, 
(d) transverse $p_{x}$-$p_{y}$ orbital $\chi^{c}_{x,y,x,y}$ 
and the pairing interactions for 
(e) $d_{x^2-y^2}(\equiv 1)$ diagonal element $V_{1111}$ 
and (f) $d_{xz}$-$d_{yz}$ off-diagonal element $V_{xz,yz,yz,xz}$ as functions of $\bm{q}$ with $i\omega_n=0$ for $f_d=0.37$ and $V'=0.345$eV with $\lambda=1$. 
}\label{Fig4}
\end{center}
\end{figure}

The $\bm{q}$ dependence of several susceptibilities and pairing interactions with $i\omega_n=0$ are shown in Fig. \ref{Fig4} 
for the same parameters in Fig. \ref{Fig3}. 
The spin susceptibility is enhanced by $d$-$d$ interaction for $\bm{q}\sim \bm{Q}_{\rm AF}=(\pi,\pi)$ due to the nesting effect (Fig. \ref{Fig4} (a)), 
while the transverse $d_{xz}$-$d_{yz}$ orbital susceptibility (Fig. \ref{Fig4} (c)) contributing to the $O_{xy}$ quadrupole one (Fig. \ref{Fig4} (b)) 
and the transverse $p_{x}$-$p_{y}$ orbital one (Fig. \ref{Fig4} (d)) are simultaneously enhanced by the $d$-$p$ OP interaction $V'$ for $\bm{q}\sim \bm{0}$ 
due to the (intersite) $\bm{q}$ dependent factor $\phi_{\alpha\beta}(\bm{q})$. 
As seen from eq. (\ref{eq:pair}) for the pairing interaction $\hat{V}(q)$, 
the FO fluctuation induces a large attractive $\hat{V}(q)$ for $\bm{q}\sim \bm{0}$ (Fig. \ref{Fig4} (f)) 
which mediates the $s$-wave pairing 
within each of the electron and the hole FSs almost independently of each other 
resulting in nearly degenerate $s_{++}$ and $s_{\pm}$-wave pairings, 
while the AFM fluctuation induces a repulsive $\hat{V}(q)$ for $\bm{q}\sim \bm{Q}_{\rm AF}$ (Fig. \ref{Fig4} (e)) 
which causes pair scattering between the electron and hole FSs and enhances the gap function with the sign change between the FSs, 
i.e., the $s_{\pm}$-wave pairing as shown in Fig. \ref{Fig3}. 
In this case, the FO and the AFM fluctuations cooperatively enhance $T_c$ without any competition by virtue of the $\bm{q}$-space segregation, 
although $T_c$ is not explicitly shown here but has the same tendency as $\lambda$ 
that is a monotonically increasing function of $f_d$ and $V'$ as shown in Fig. \ref{Fig2} (b).

In summary, we have found that 
the $d$-$p$ OP interaction derived from the orbital dependence of the Coulomb integrals between Fe $d$ and As $p$ orbitals 
is crucial for enhancing the FO fluctuation responsible for the $C_{66}$ softening. 
The FO fluctuation induces the attractive pairing interaction for $\bm{q}\sim \bm{0}$ 
which mediates the $s$-wave pairing within each of the electron and the hole FSs, 
while the AFM fluctuation enhanced by the $d$-$d$ interaction induces the repulsive one for $\bm{q}\sim \bm{Q}_{\rm AF}$ resulting in the $s_{\pm}$-wave pairing, 
where the experimentally observed FO and AFM fluctuations cooperatively enhance $T_c$ without any competition by virtue of the $\bm{q}$-space segregation. 

In the present study, we have focused only on the $d_{x'z}$-$d_{y'z}$ ($p_{x'}$-$p_{y'}$) type OP interaction responsible for the $C_{66}$ softening, 
but a preliminary examination with including the other types shows 
that the $d_{xz(yz)}$-$d_{x^2-y^2}$ ($p_{z}$-$p_{x(y)}$) type OP interaction enhances the $O_{3z^2-r^2}$ quadrupole fluctuation responsible for the $C_{33}$ softening 
which was recently observed by the ultrasonic experiment\cite{JPSJ.82.114604} and was discussed by the mode-coupling theory\cite{arXiv:1312.0481}. 
More recently, an electron diffraction experiment revealed the strong coupling of Fe and As orbital polarizations along the $c$-axis\cite{arXiv.1401.3706} 
which can be also explained by the $d$-$p$ OP interaction between Fe $d_{xz(yz)}$-$d_{x^2-y^2}$ and As $p_{z}$-$p_{x(y)}$ polarizations. 
Hence, detailed calculations with including the complete matrix elements of the $d$-$p$ interaction together with the electron-phonon interaction 
crucial for explicit description of the phonon softening\cite{C66,phonon} are important future problems.

Finally, we briefly discuss the impurity effect on $T_c$ which is robust against nonmagnetic impurities\cite{JPSJ.79.014710} 
and is not consistent with the $s_{\pm}$-wave pairing\cite{PhysRevLett.103.177001}. 
As for the $s_{\pm}$-wave pairing obtained in the present study, 
however, the impurity effect is considered to largely depend on the parameters: 
it is large for the AFM fluctuation-dominated case 
where $T_c$ is mainly determined by the repulsive pairing interaction for $\bm{q}\sim \bm{Q}_{\rm AF}$ 
while small for the FO fluctuation-dominated case 
where $T_c$ is mainly determined by the attractive one for $\bm{q}\sim \bm{0}$. 
Explicit calculations of the impurity effect on $T_c$ are now under way and will be reported in a subsequent paper. 

\begin{acknowledgment}
The authors thank Y. Yanagi for useful comments and discussions. 
This work was partially supported by a Grant-in-Aid for Scientific Research from 
the Ministry of Education, Culture, Sports, Science and Technology, 
and also by a Grant-in-Aid for JSPS Fellows. 
\end{acknowledgment}

\bibliography{fe_dp}

\begin{thebibliography}{10}
\bibitem{C66}
A comparison between the phonon Green's function $D(q)$ and the quadrupole
  susceptibility $\chi_O(q)$ yields $C_{66}/C_{66}^0=\lim_{\bm{q}\to
  \bm{0}}[\omega(\bm{q})/\omega^0(\bm{q})]^2= \chi_{O_{xy}}^{\rm irr}(0)/
  \chi_{O_{xy}}(0)$, where $\omega(\bm{q})(\omega^0(\bm{q}))$ is the
  renormalized (bare) frequency of the acoustic phonon coupled to the
  quadrupole $O_{xy}$ and $\chi_{O}^{\rm irr}$ is the contribution from the
  irreducible diagrams with respect to $D(q)$. When approaching the structural
  transition with $C_{66}\to 0$, the system also shows the FO instability with
  $\chi_{O_{xy}}(0) \to \infty$.
\bibitem{phonon}
The frequency $\omega(\bm{q})$ of a phonon coupled to a quadrupole $O$ is
  renormalized as $\omega(\bm{q})/\omega^0(\bm{q})= [\chi_{O}^{\rm
  irr}(\bm{q},0)/ \chi_{O}(\bm{q},0)]^{1/2}$ and is expected to show a
  softening for $\bm{q}\sim \bm{Q}_{\rm AF}$ due to the enhancement of the AFO
  fluctuation as observed for the acoustic phonon softened for $\bm{q}\sim
  \bm{0}$ due to the FO fluctuation\cite{C66}.
\end{thebibliography}
\end{document}